\documentclass[referee]{raa}            

\usepackage{booktabs,threeparttable}
\usepackage{multirow}
\usepackage{longtable}
\usepackage{graphicx,times}             
\input{epsf.sty}                        
\input{psfig.sty}                       

\begin{document}

   \title{Delayed Onset and Fast Rise of Prompt Optical-UV Emission from
Gamma-Ray Bursts in Molecular Clouds
}

   \volnopage{Vol.0 (200x) No.0, 000--000}      
   \setcounter{page}{1}          

   \author{Xiao-Hong Cui
      \inst{1}
      Zhuo Li
      \inst{2,3}
      \and Li-Ping Xin
      \inst{1}
        }

   \institute{National Astronomical Observatories, Chinese Academy of Sciences,
             Beijing 100012, China; {\it xhcui@bao.ac.cn}\\
             \and
             Department of Astronomy, Peking University, Beijing 100871, China\\
                \and
             Kavli Institute for Astronomy and Astrophysics, Peking
University, Beijing 100871, China\\
   }

   \date{Received~~2009 month day; accepted~~2009~~month day}

\abstract{Observations imply that long $\gamma$-ray bursts (GRBs) are
originated from explosions of massive stars, therefore they may
occur in the molecular clouds where their progenitors were born.  We
show here that the prompt optical-UV emission from GRBs may be
delayed due to the dust extinction, which can well explain the
observed optical delayed onset and fast rise in GRB 080319B. The
density and the size of the molecular cloud around GRB 080319B are
roughly constrained to be $\sim10^3$cm$^{-3}$ and $\sim 8$pc,
respectively. We also investigate the other GRBs with prompt
optical-UV data, and find similar values of the densities and sizes
of the local molecular clouds. The future observations of prompt
optical-UV emission from GRBs in subsecond timescale, e.g., by
UFFO-Pathfinder and SVOM-GWAC, will provide more evidence and probes
of the local GRB environments.  }

   \authorrunning{X.-H. Cui \& Z. Li }            
   \titlerunning{Prompt Optical/UV Emission from GRBs}  

   \maketitle

%
%
\section{Introduction}           
\label{sect:intro}

The properties of $\gamma$-ray burst (GRB) circumburst and
host-galaxy environment are important for the studies of GRB
progenitors and the fundamental conditions required within a galaxy
to form a GRB. The multi-wavelength observations about the emission
from GRBs and that from their host galaxies would provide a unique
tool to understand the nature of GRBs and the properties of
interstellar medium (ISM) around the bursts.

Observations imply that long GRBs are originated from massive star
explosions. First, they are observed to lie in star-forming
galaxies, or even within the active star-forming regions of the host
galaxies (e.g., Paczy\'{n}ski 1998; Bloom, Djorgovski, \& Kulkarni
2002). More precise HST images of
afterglows reveals that they occur within a few kiloparsecs of the
flux-weighted centroid of their host galaxies (Fruchter et
al. 2006). Second, X-ray observations show evidence for high column
densities of gas around long GRBs, implying giant molecular clouds
around them (e.g., Galama \& Wijers 2001). Finally, at least some
long GRBs are associated with core-collapse supernovae (SNe). The
discovery of four clear associations between long, soft GRBs and
Type Ib/c SNe and many SN-like bumps in the late optical afterglow
light curves (see, e.g., review by Woosley \& Bloom 2006) directly
tell that their progenitors are massive stars.

Now that the progenitors of long GRBs are massive stars, they may
occur in the birth place of the progenitors since massive stars are
short-lived, i.e., the long GRBs may lie in the molecular clouds
that the massive stars are born. The optical-UV and X-ray emission
from GRBs can be affected significantly by the extinction of dust
and absorption of gas in the local environment, $\gamma$-ray
emission is almost unaffected. Therefore, one may naturally expect
that the behavior of prompt optical-UV emission is different in
light curves from that of prompt $\gamma$-ray emission. The
difference may hint the properties of the dust environments around
the GRBs. The interaction of a GRB with the environment can yield powerful clues on the properties of the medium in which the burst occur. The behavior of the X-ray and optical opacities in the nearby of a GRB have been studied (Perna, Raymond, \& Loeb 2000; Perna\& Raymond 2000). A time-dependent photoionization code has been developed to study the modificaitons in the dust distribution and the graphite in the medium around the GRB was found to be more resistent than silicates (Perna \& Lazzati 2002; Lazzati \& Perna 2002).

The varieties of observed GRB prompt optical behaviors are rich. The
prompt optical emission was first observed in GRB 990123 and was
found to be uncorrelated with the ongoing $\gamma$-ray emission
(Akerlof et al. 1999; but see Liang et al. 1999). Then the prompt
optical emission from GRB 050820A (Vestrand et al. 2006) was
reported and a strong correlation between $\gamma$-energy and
optical emission in the prompt phase was discovered. Similar cases of some
degree of correlation are observed in GRB 041219A (Vestrand et al.
2005; Blake et al. 2005), GRB 060526 (Th\"{o}ne et al. 2010), and
``naked eye'' burst GRB 080319B (Racusin et al. 2008; Beskin et al.
2010). GRB 080319B with the richest prompt optical observation
attracted much attention about it's nature. The detailed observation
of this burst presented by Racusin et al. (2008) showed not only a
correlation between $\gamma$-ray and optical emission in the prompt
phase but also an obvious delayed onset $\sim$ 15s between them.

In this work, we show that if a GRB is located in a molecular cloud,
its prompt optical-UV emission may be absorbed by the dust in the
molecular cloud, and only emerges after the dusts on the line of
sight are all destroyed. This can well explain the observed delayed
onset of the prompt optical-UV emission in GRB 080319B, and the
density and the size of the molecular cloud around this burst can be
roughly constrained. For other bursts with prompt optical
observations, the properties of local environment can also be
constrained. We find similar properties of the clouds, with density
and size being $n_{\rm H}\sim 10^3-10^4$ cm$^{-3}$ and $\Delta R\sim 6$
pc. The paper is arranged as the following: a simple model of the
radiation-dust interaction and the resulted prompt optical-UV light
curve are presented in $\S$ 2; in $\S$ 3, we apply the model to GRB
080319B and other GRBs with prompt optical-UV observations; in $\S$
4 discussions and conclusion are presented.

\section{Radiation-dust interaction and emergent optical-UV emission}

Consider a GRB that is located in a molecular cloud. The prompt
optical-UV emission from this GRB may be absorbed by the dust in the
cloud, but in the same time the dusts may also be destroyed by the
emission. If the optical-UV emission is strong and lasts long
enough, it may emerge from the cloud after the dusts on the way are
all destroyed. The dust destruction by the optical-UV radiation has
been discussed by Waxman \& Draine (2000). Here we will follow their
model in the radiation-dust interaction, and focus on the back
effect of the dust on the optical-UV emission, i.e., how the dusts
in a cloud of finite size affects the apparent light curve of the
prompt optical-UV emission. On the other hand, from the observed
light curve profile of the prompt optical-UV emission, we can also
give some constraints on the properties of the molecular cloud. We
will only consider dust destruction due to thermal sublimation and
neglect the effect of grain charging, since, as argued by Waxman \&
Draine (2000) and Draine \& Hao (2002), the thermal sublimation is
likely to be more effective (see, but, Fruchter et al. 2001).

\begin{figure*}
\includegraphics[angle=0,scale=0.4]{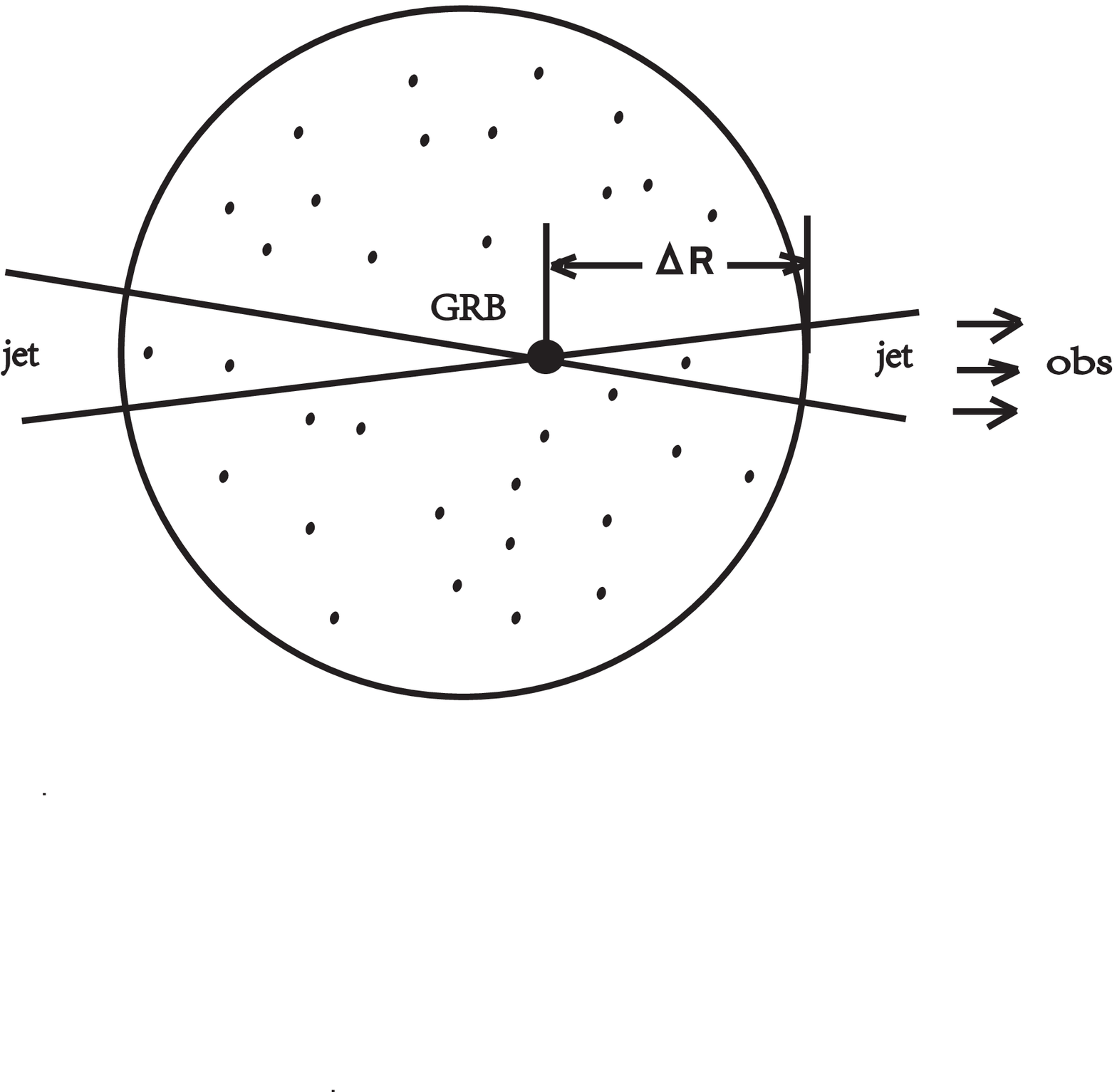}
\hfill \caption{The sketch of an observed GRB located inside a
molecular cloud. "obs" denotes the direction to the Earth
(observer). The cloud is assumed to be with uniform density and a
clear boundary. The distance of the GRB to the edge of the cloud is
$\Delta R$, as marked. \label{sketch}}
\end{figure*}

Considering a simple picture as shown in Fig.\ref{sketch}, the
molecular cloud is assumed to be uniform in density, and the
distance of the GRB from the edge of the cloud on the side to the
observer is $\Delta R$. The cloud contains dust grains of
characteristic radius $a$ and dust number density $n_d$. Assuming a
standard dust-to-gas mass ratio, $n_d$ is related to the cloud
density $n_{\rm H}$ as $n_d=0.01n_{\rm H}m_{\rm H}/(4\pi/3)a^3\rho$,
where $\rho$ is mass density of the grain material. A characteristic
value of $\rho = 3.5$g cm$^3$ (Guhathakurta \& Draine 1989) will be
taken in the following calculations. Considering that the source
radiates a $1-7.5$~eV radiation with luminosity $L_{1-7.5}$, a
grain at a distance $r$ can be heated up leading to thermal
sublimation and thermal emission. The temperature $T$ of the grain
at distance $r$ from the source is governed by
\begin{equation}
  \frac{L_{1-7.5}}{4\pi r^2}Q_{\rm UV}\pi a^2=\langle Q\rangle_T4\pi a^2\sigma T^4-4\pi a^2{da\over
  dt}{\rho\over m}B,
\end{equation}
where $m$ is the mean atomic mass, $B$ is the chemical binding
energy per atom, $Q_{\rm UV}$ is the absorption efficiency factor
averaged over the $1-7.5$~eV spectrum of the source emission, and
$\langle Q\rangle_T$ is the usual Planck-averaged absorption
efficiency. We will assume $Q_{\rm UV}\approx1$ for $a\ga10^{-5}$cm
and approximate $\langle Q\rangle_T$ by
\begin{equation}
  \langle Q\rangle_T\approx\frac{0.1a_{-5}(T/2300~\rm K)}{1+0.1a_{-5}(T/2003~\rm K)}
\end{equation}
with $a_{-5}=a/10^{-5}\rm cm$. The thermal sublimation rate can be
approximated by (Guhathakurta \& Draine 1989)
\begin{equation}
  {da\over dt}=-(\frac{m}{\rho})^{1/3}\nu_0e^{-B/kT}.
\end{equation}
We adopt the frequency $\nu_0=1\times10^{15}\rm s^{-1}$, $B/k=7\times10^4$K, and
$\rho/m=10^{23}\rm cm^{-3}$ as representative values (Guhathakurta \& Draine 1989; Waxman \&
Draine 2000). If we assume
$T$ is approximately constant during the illumination, then the
grain survival time at $T$ is $t_{\rm surv}(T)=a/|da/dt|$. The grain
will be completely destructed by thermal sublimation if it is
illuminated over a time longer than $t_{\rm surv}(T)$.

As the dust is destructed by the radiation, the radiation is also
extinguished by the dust. We consider below the effects of dust
extinction on the observed flash light curves. Following Waxman \&
Draine (2000) let us approximate the $1-7.5$ eV emission from the
GRB as a rectangular pulse of duration $\Delta t$ and luminosity
$L_{1-7.5}$. The problem can be simplified by assuming that the
effects of extinction can be approximated as a narrowing of the
optical pulse, retaining a rectangular profile. The leading edge of
the radiation is just at the dust destruction front. We assume a
sharp disruption front within which the dust grains are all
destructed whereas the grains further are not affected. Define
$f(r)$ the fraction of the flash energy that is absorbed by dust
interior to radius $r$. If $t_{\rm surv}<(1-f)\Delta t$, the grains
are destructed and $f(r)$ satisfies
\begin{equation}
  \frac{df}{dr}=Q_{\rm UV}n_d\pi a^2\frac{t_{\rm surv}}{\Delta t}.
\end{equation}
The relation between the radius of dust destruction front $R_f$ and
observer time $t_{\rm obs}$  can be given by
\begin{equation}
  t_{\rm obs}=f(R_f)\Delta t(1+z),
\end{equation}
with $z$ the redshift of the GRB source. A (maximum) dust
destruction radius $R_d$ is determined by the condition $t_{\rm
surv}[T(R_d)]=[1-f(R_d)]\Delta t$. At $r>R_d$, the dust grain
survives the illumination and the destruction front does not move
any more, therefore we can simply assume $R_f=R_d$ and $f=1$ at
$t_{\rm obs}>f(R_d)\Delta t(1+z)$.

The above discussion on $R_f$ propagation omits the existence of the
edge of the cloud at $\Delta R$. If $R_f<\Delta R$, the optical
depth due to dust extinction is
\begin{equation}
  \tau=Q_{\rm UV}n_d\pi a^2(\Delta R-R_f),
\end{equation}
and the attenuated luminosity observed outside is
\begin{equation}\label{Luminosity}
  L_{\rm obs}(t_{\rm obs})=L_{1-7.5}\exp\{-\tau[R_f(t_{\rm obs})]\}.
\end{equation}

If $R_d>\Delta R$ then the destruction front can reach the edge
($R_f=\Delta R$) at time $t_{\rm obs}=f(\Delta R)\Delta t(1+z)$,
which means all the dust in the beam of the radiation is cleared and
the radiation is not attenuated, $\tau=0$. However, if $R_d<\Delta
R$ then the dust is not destructed completely, and the disruption
front stays at $R_d$ at $t_{\rm obs}>f(R_d)\Delta t(1+z)$ while the
dust optical depth is fixed at $\tau=Q_{\rm UV}n_d\pi a^2(\Delta
R-R_d)$.

Note, in the former case $R_d>\Delta R$, the $1-7.5$ eV emission is
firstly totally attenuated, since $R_f<\Delta R$ and $\tau\gg1$;
when the destruction front propagates to be close to the edge of the
cloud, $R_f\la\Delta R$ and $\tau\sim1$, it starts to emerge by some
fraction; and after the destruction front reaches the edge,
$R_f=\Delta R$, the emission emerge completely without any
extinction. Thus, the end time of the light curve rising up depends
on the edge of cloud $\Delta R$, while the slope of the light curve
rising up depends on the propagation speed of the destruction front
which is sensitive to the cloud density, $n_{\rm H}$.

\section{Applications}

As discussed above, in the case of $R_d>\Delta R$, the
radiation-dust interaction leads to that only the later part of the
prompt optical-UV photons emerges, but the $\gamma$-ray photons from
the GRB are without any attenuation. Thus, if the prompt
$\gamma$-ray and optical-UV radiation is emitted together from the
GRB source, there should be a time delay between the onset of the
apparent prompt optical-UV and $\gamma$-ray emission. So far there
are quite a few GRBs that are detected with prompt optical-UV
emission. We will apply the simple radiation-dust interaction model
to all these detected GRBs, with the goal to explain the time delays
of the prompt optical-UV emission relative to $\gamma$-ray emission,
and roughly give some implications to the properties, e.g., the
densities and the sizes of the molecular clouds around them.

The observed luminosity is usually given  in a single band for a
single filter, e.g., U, B, V, R bands etc. A spectrum with the form
$f_\nu \propto \nu^{-1}$ is assumed for the prompt optical-UV flash
in 1--7.5 eV, which is consistent with the fast-cooling electrons
expected in the standard internal shock model. Thus, a cosmological
$\kappa$--correction factor can be defined to account for the
transformation of the single passband of filter to the band of
1--7.5 eV in the proper GRB frame,
\begin{equation}
 \kappa=\frac{ \int ^{7.5{\rm eV}/h(1+z)} _{1{\rm eV}/h(1+z)} f_\nu d
 \nu}{ \int ^{b_2} _{b_1} f_\nu d\nu}
\end{equation}
where $b_2$ and $b_1$ are the frequency boundaries of the passband
for the observed filter, $z$ is the GRB redshift.

\subsection{GRB 080319B}
So far the so-called ``naked-eye'' GRB 080319B is the only one that
happens to occur in the field of view of an optical telescope,
without the trigger by high-energy detector, thus it is by luck
monitored in optical band from before the beginning of the GRB. The
broadband observations of it has been presented by Racusin et al.
(2008) and Beskin et al. (2010). The $\gamma$-ray emission was found
to begin at about 4 s before the BAT trigger and last $\sim$ 57 s.
The bright optical transient begins at $\sim$ 10 s after the BAT
trigger, peaks at $\sim$18 s and then fades below the threshold to
magnitude $\sim$12 after 5 min. That is to say, there is a time
delay $\sim$ 14 s between the onsets of $\gamma$-ray and optical
emission. It should be noticed that the optical rising is too fast
to be accounted for by the afterglow model, either forward shock
emission (Sari et al. 1998) or reverse shock emission (Kobayashi et
al. 2000). The optical light curve during the plateau phase shows
fluctuation, similar to the $\gamma$-ray one. Moreover, the optical
and $\gamma$-ray emission is found to be correlated. All these
features suggest that the prompt optical emission from this burst is
not produced by afterglow shock. Thus the delayed up-rising optical
emission needs other explanation. We show below that the delay can
be well explained by the radiation-dust interaction.

We, again, approximate the intrinsic optical-UV emission as a
rectangle pulse of duration $\Delta t$ (in the rest frame of the
GRB). Since the optical emission is observed to decay at
$\approx50$s (similar to the $\gamma$-ray duration), the duration is
$\Delta t\approx50/(1+z)\approx25$s, where the GRB redshift is $z =
0.937$ (Vreeswijk et al. 2008). Apparently in observations, the
optical flux rises from zero to a plateau phase at a time
$t_b\approx 15$s after trigger, and the plateau phase ends at $\sim$
50 s. The mean luminosity after $t_b$ (i.e.,in the range of 15--50
s) and in 1--7.5 eV energy band can be given by $L_{1-7.5} = 4\pi
D_L(z)^2 \kappa f_p \approx 2.6 \times 10^{50}$ erg s$^{-1}$. Here
$D_L(z)$ is the luminosity distance calculated (adopting a Universe
model with $\Omega_\mathrm{M}=0.3$, $\Omega_{\Lambda}=0.7$, and
$H_0=71$ km $\rm{s}^{-1}$ $\rm{Mpc}^{-1}$), and $f_p=9.39 \times 10^{-9}$ erg cm$^-2$ s$^{-1}$ is the mean
flux observed during the time range 15--50 s in V band observed by
TORTORA (Pagani et al. 2008; Racusin et al. 2008). The correction factor $\kappa=6.17$ for this burst is
calculated by assuming a power law spectrum $f_\nu \sim \nu^{-1}$.

\begin{figure*}
\includegraphics[angle=0,scale=0.40]{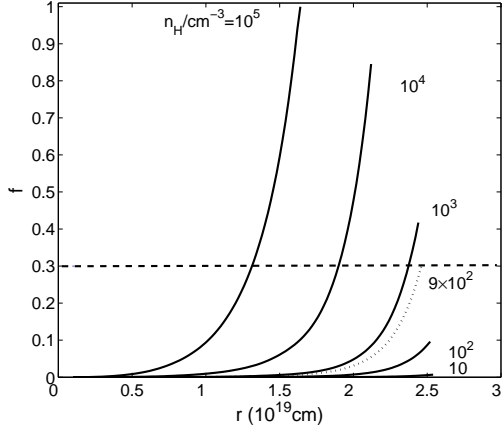}
\hfill \caption{The fraction $f(r)$ of flash energy absorbed by dust
interior to radius $r$ up to dust destruction radius $r = R_d$ in
the case of GRB 080319B. The duration and luminosity of prompt
emission in 1--7.5 eV are $\Delta t = 25$ s and $L_{1-7.5} = 2.6
\times 10^{50}$ erg s$^{-1}$, respectively, and the dust grain size
is assumed to be $a = 10^{-5}$ cm. Different lines correspond to
different values of cloud density $n_{\rm H}$, as marked in the
plot. The dash line shows $f (r = \Delta R) = 0.3$. The dotted line
presents the case of $f(r=R_d)=0.3$, i.e., $R_d=\Delta R$.
\label{fr}}
\end{figure*}

If the duration of the optical-UV emission $\Delta t = 25$ s is
taken and assuming the radius of dust grain as $a = 1\times 10^{-5}$
cm, we calculate the absorbed energy fraction $f(r)$ of the flash up
to the destruction radius $R_d$ for a cloud density range of $n_{\rm
H} = 10^2 - 10^5$cm$^{-3}$. The result is shown in Fig.\ref{fr}. We
can see the denser the cloud the faster the flash energy is
absorbed. However, after the destruction front reaches the
destruction radius $r=R_d$, the absorbed fraction rapidly reaches
unity, $f(R_d)=1$.

Apparently in observations, the optical emission rises to the mean
flux level at about 15 s, and then keeps this level until $\sim$ 50
s. This implies that the absorbed fraction of the flash energy, when
the dust destruction front reaches the edge of the cloud, is $f (R_f
= \Delta R)\approx 15/50 = 0.3$. Thus, given the cloud density
$n_{\rm H}$, the cloud size, roughly implicated by $\Delta R$, can
be determined for this burst, i.e. $n_{\rm H}$ and $\Delta R$ are
one by one related for fixed $f(r=\Delta R)$ value. For example, if
$n_{\rm H} =(10^3, 10^4, 10^5)$cm$^{-3}$, we have $\Delta R = (2.4,
1.7, 1.1)\times 10^{19}$cm, respectively with fixed $f (r=\Delta R)
= 0.3$. Then the value of $\Delta R$ can be found to decrease with
larger value of $n_{\rm H}$. However, for too small $n_{\rm H}$, the
destruction front reaches the maximum destruction radius $R_d$
before reaching the edge of the cloud, i.e., the absorbed fraction
$f(r=R_d)<0.3$, as the case of $n_{\rm H}=10^2$cm$^{-3}$ in Fig
\ref{fr}.

\begin{figure*}
\includegraphics[angle=0,scale=0.45]{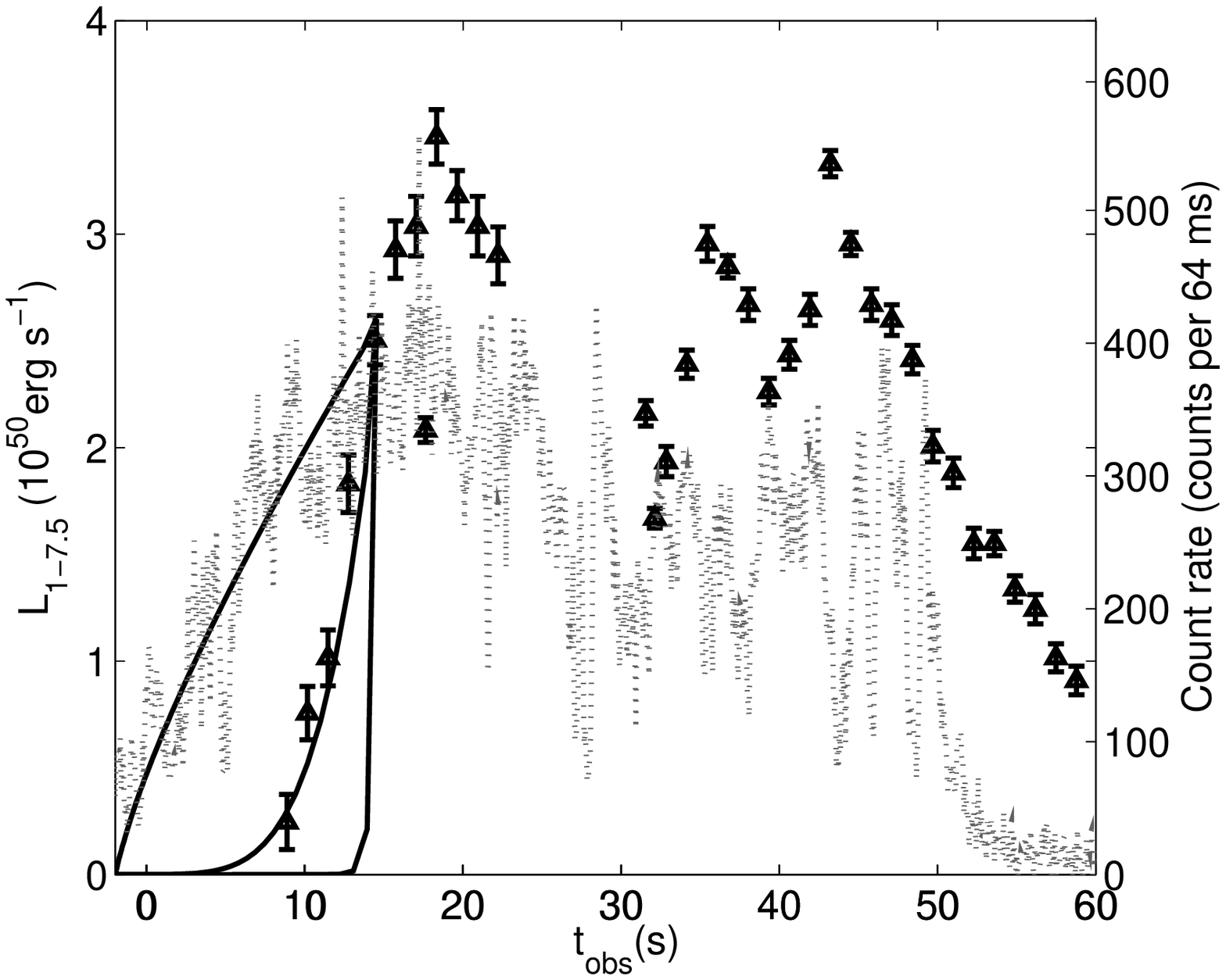}
\includegraphics[angle=0,scale=0.45]{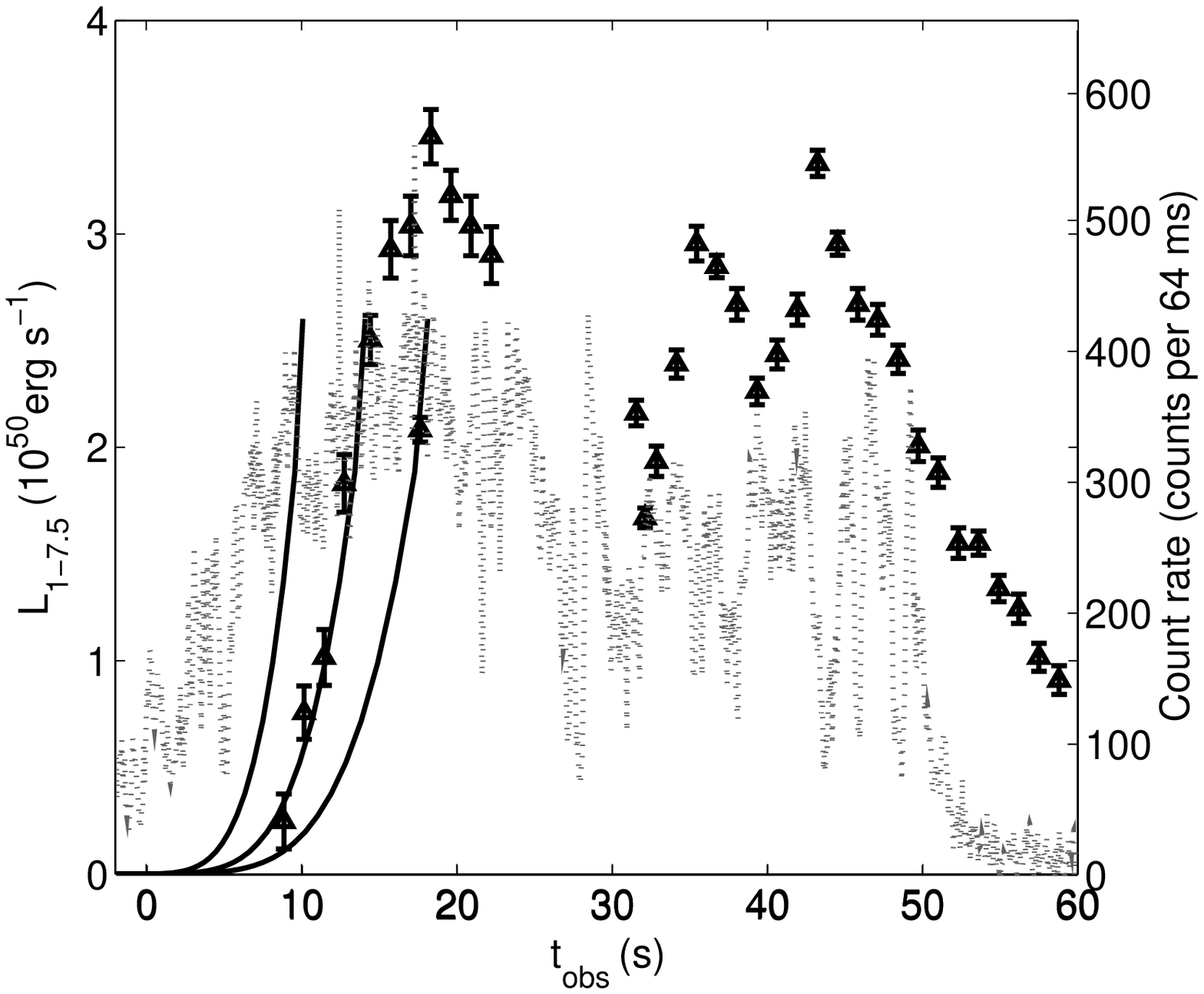}
\hfill \caption{The light curves of GRB 080319B in $\gamma$-ray and
optical bands. The black triangles are optical data from TORTORA.
For comparison, the Konus-Wind background subtracted $\gamma$-ray
light curve (18--1,160 keV), with respect with the trigger time by
Swift-BAT, is shown with dashed line. The solid lines are the
calculated optical light curves in the simple radiation-dust
interaction model. Left panel: The cases with the cloud density
$n_{\rm H}=10^2, 10^3, 10^4$cm$^{-3}$ (corresponding to three light
curves from left to right respectively) and the fixed end time of
the rising part $t_b=16$ s. Right panel: The cases with $t_b = 12,
16, 20$ s (from left to right) and the fixed $n_{\rm
H}=10^3$cm$^{-3}$. The other parameters are the same as
Fig.\ref{fr}. The case with $t_b=16$ s and $n_{\rm H}=10^3$cm$^{-3}$
gives the best fit to the rising part of the optical flash of GRB
080319B. \label{lightcurve}}
\end{figure*}

In order to decouple $n_{\rm H}$ and $\Delta R$, we need to further
consider the temporal profile of the observed optical-UV emission.
For different values of $n_{\rm H}$ and $t_b$, we have calculated
the optical-UV light curve using equation (\ref{Luminosity}). The
resulted light curves are shown in Fig.\ref{lightcurve}, also
plotted are the $\gamma$-ray and optical-UV data that are adopted
from Racusin et al. (2008). Note, as Beskin et al. (2010) found that
the optical emission is 2 s delayed relative to the $\gamma$-ray
emission in the plateau phase, we also assume a time delay of 2 s
for the intrinsic onset of optical-UV emission compared to the
$\gamma$-ray one. The plotted light curves in Fig. \ref{lightcurve}
take this into account. We see that compared with the observed
optical data of GRB 080319B, the case with $n_{\rm H} =
10^3$cm$^{-3}$ and $t_b=16$ s fits the light curve profile better.
Therefore, it can be concluded that the cloud that hosts GRB 080319B
has a density of $n_{\rm H} \approx 10^3$cm$^{-3}$ and a size of
$R\sim\Delta R \approx 8$ pc.

It should be noted that in the above calculations we have taken
$\rho = 3.5$g cm$^3$ (Guhathakurta \& Draine 1989), $a = 1\times
10^{-5}$cm and a standard dust-to-gas mass ratio of 0.01. The
resulted values of $n_{\rm H}$, $\Delta R$ and $R_d$ are not
sensitive to the values of them, i.e., the resulted $n_{\rm
H}$, $\Delta R$ and $R_d$ values vary within a factor of a few if
changing $\rho$, $a$, $\kappa$, and the gas-to-dust ratio by one
order of magnitude. This is good enough for order of magnitude
estimate with the simple model here.

\subsection{Other GRBs with prompt optical detections}
Besides GRB 080319B, there are quite a few other GRBs with prompt
optical detections during the $\gamma$-ray bursting phase. They are
all detected by rapid slew of optical telescopes to the GRB location
after trigger by $\gamma$-ray detectors. So usually there is a gap
between the trigger time and the start time of optical observation.
Nevertheless, we can still try to make some constraints on the local
GRB environments based on the simple radiation-dust interaction
model.

All the GRBs detected after December of 2004 and with optical
detections during the prompt $\gamma$-ray emission are collected and
analyzed with the simple radiation-dust interaction model here. We
separate these GRBs into two samples. In {\em Sample I}, the GRBs
satisfie the following three criterions: (1) There are optical
detections before the end of the GRB, specifically, the optical
detection is within the duration of $T_{90}$; (2) The optical light
curve within $T_{90}$ shows a rising of the flux, i.e., if the
optical light curve shows a decay or flat plateau then the GRB is
not included; (3) The number of optical data points, excluding upper
limits, in the rising part is not less than three. All the other
GRBs only satisfy criterion (1) are grouped into {\em Sample II}.

We find besides GRB 080319B, there are 7 other GRBs satisfy the
three criterions: GRB 041219A (Vestrand et al. 2005; Blake et al.
2005); GRB050820A (Vestrand et al. 2006); GRB 060218 (Mirabal et al.
2006; Ferrero et al. 2006; Sollerman et al. 2006; Kocevski et al.
2007); GRB 060418 (Molinari1 et al. 2007; Dupree et al. 2006;
Vreeswijk \& Jaunsen 2006), GRB 060607A (Molinari1 et al. 2007;
Ledoux et al. 2006), GRB 080810 (Page et al. 2009; Burenin et al.
2008), and GRB 100906A (Gorbovskoy et al. 2011; Barthelmy et al.
2010; Markwardt et al. 2010; Tanvir et al. 2010). However, we
exclude GRB 041219A and GRB 060218 from Sample I for reasons as
follows. GRB 041219A shows correlation between $\gamma$-ray and
optical emission, thus the observed initial rising in the optical
band is likely intrinsic (Vestrand et al. 2005) other than due to
radiation-dust interaction. This GRB is included in Sample II
instead. As for GRB 060218, its early optical-UV emission is likely
associated with the supernova shock breakout (Campana et al. 2006;
Waxman et al 2007), thus not due to radiation-dust interaction
either. Sample I GRBs are listed in Table \ref{Tab:poptical}. All
the other GRBs only satisfy criterion (1) are grouped into {\em
Sample II}. For example, GRB 110205A (Klotz et al. 2011a, b;
Cucchiara et al. 2011) was started to be detected in optical band
166 s after trigger but within the duration of $T_{90}=257$ s. The
observed optical light curve already appears to be a plateau,
without a rising part, which may occur before the start of optical
detections. We exclude this burst in Sample I but include in Sample
II. As shown in Table \ref{Tab:limit}, there are 13 GRBs in Sample
II.

\begin{table*}
  \caption[]{The observational results of GRBs in Sample I and the constraints of their local molecular clouds}
  \label{Tab:poptical}
  \begin{center}\begin{tabular}{ccccccccccccc}
  \hline\noalign{\smallskip}
GRB &  $z$  & $T_{90}$   & $t_{\rm op}$ &$\kappa^*$ &  $L_{1-7.5}$ &$R_{\rm d}$& $\Delta R$ & $n_{\rm H}$ & $\Delta t_{\rm obs}$ & $t_b$& ref  \\
&  & (s)  & (s) & &  ($10^{48}$ erg s$^{-1}$) &(pc)& (pc) & ($10^3$ cm$^{-3}$)& (s) & (s)  & \\
  \hline
  050820A &2.6& ~750 $^{**}$ & 84 & 3.10 (R) & 0.6 &3.35 & 3.23 & 9 & 646&305 & 1, 2, 3 \\
060418 & 1.49 & 103.1 & 40 &4.07 (H)  &2.4 & 6.91 & 6.87 &4&140 &107&4, 5, 6 \\
060607A & 3.082 & 102 & 73 &4.07 (H)  & 2.3 & 5.90& 5.87 & 5& 200  &150& 4, 7 \\
080319B & 0.937 & 57 &8.9 $^{\dag}$ & 6.17 (V) & 260 & 7.78& 7.67& 1& 50 & 16& 8, 9, 10  \\
080810 & 3.35& 106 &38& 3.10 (W) & 7.1 & 11.7 & 11.4 &3  & 150 & 67& 11, 12 \\
100906A & 1.727 & 114.4 & 48.5 & 3.10 (W) & 1.2 & 4.60 & 4.48 & 15& 190 & 83&13, 14, 15, 16 \\
\hline
  \end{tabular}\end{center}
  \begin{tablenotes}
  \item[*]* In the bracket is the passband of filter. Letters ``V'', ``H'' and ``W'' denote V, H and white bands, respectively.
  \item[*]** From the work of Vestrand et al. (2006), rather than Swift data.
  \item[*] $\dag$ The time corresponds to the first optical data by TORTORA. In fact, the optical observations start before the trigger of this GRB.
  \item[*]References: (1) Prochaska et al. 2005; (2) Ledoux et al. 2005; (3) Vestrand et al. 2006; (4) Molinari1 et al. 2007; (5) Dupree et al. 2006; (6) Vreeswijk \& Jaunsen 2006; (7) Ledoux et al. 2006; (8) Vreeswijk et al. 2008; (9) Racusin et al. 2008; (10) Beskin et al. 2010; (11) Page et al. 2009; (12) Burenin et al. 2008; (13) Gorbovskoy et al. 2011; (14) Barthelmy et al., 2010; (15) Markwardt et al., 2010; (16) Tanvir et al. 2010
 \end{tablenotes}
\end{table*}

\begin{table*}
  \caption[]{The observational results of GRBs in sample II and the constraints of their local molecular clouds}
  \label{Tab:limit}
  \begin{center}\begin{tabular}{ccccccccccccc}
  \hline\noalign{\smallskip}
GRB &  $z$  & $T_{90}$   & $t_{\rm op}$  &   $L_{1-7.5}$  & ref  \\
&  & (s)  & (s) &   (10$^{47}$erg s$^{-1}$) &   \\
  \hline\noalign{\smallskip}
041219A& 0.31 & 520& 460&3.7$\times 10^{-5}$ & 1,2\\
050319 & 3.24  &160.5 & 30.4 & 3.9 & 3,4   \\
050904 & 6.29 & 174.2 & 150.3 & 1.1 &5    \\
060526 & 3.21 & 298.2 & 16.1 & 3.4  &6 \\
060904B & 0.703 & 171.5 & 21 & $<$6.8$\times 10^{-2}$  &7 \\
061126 & 1.16& 70.8& 42 &1.9$\times 10^{-2}$ &8,9 \\
071003 & 1.1 & 150 & 44.5 & 3.0 &10 \\
071031 & 2.69 & 180 & 59.6 & 2.3 &  11,12 \\
080603A & 1.69& 150 &105 &5.8$\times 10^{-3}$  & 13\\
080607 & 3.036 & 79 & 24.5 & 2.7  & 14\\
100901A & 1.408 & 439 & 113.4 &9.6$\times 10^{-2}$ &15,16,17\\
100902A & 4.5 & 428.8 & 104 & $<$1.4 & 17\\
110205A & 1.98 & 257 & 166 & 0.2 & 18 \\
  \noalign{\smallskip}\hline
  \end{tabular}\end{center}
  \begin{tablenotes}
  \item[*]References: (1) Vestrand et al. 2005; (2) Blake et al. 2005 (3) Quimby et al. (2006); (4) Wozniak et al. 2005; (5) Bo\"er et al. 2006; (6) Th\"one et al. 2010: (7) Klotz et al. 2008; (8) Gomboc et al. 2008; (9) Perley et al. 2008a; (10) Perley et al. 2008b; (11) Kruehler et al. 2007; (12) Antonelli et al. 2007; (13) Guidorzi et al. 2011; (14) Perley et al. 2011; (15) Chornock et al. 2010; (16) Immler et al. 2010; (17) Gorbovskoy et al. 2011; (18) Cucchiara et al. 2011
 \end{tablenotes}
\end{table*}

\subsubsection{Sample I}
For GRBs in Sample I, we can follow the same approach we carry for
GRB 080319B, and the optical rising can be accounted for by the
simple radiation-dust interaction model, in the same time the local
environments of these bursts are constrained by fitting the observed
promptly optical-UV light curves.

We assume there is intrinsic optical-UV emission associated with the
$\gamma$-ray emission, with approximated rectangle light curve
profile. The optical-UV duration $\Delta t_{\rm obs}=\Delta t(1+z)$
is obtained from observations (which is usually comparable or
somewhat larger than the $\gamma$-ray duration). The luminosity
$L_{1-7.5}$ is calculated from observed optical emission, by
correction with $\kappa$ factor assuming a $f_\nu\propto\nu^{-1}$
spectrum (For white band the same $\kappa$ factor as R band is
assumed). There are usually fluctuations of optical flux in the
plateau phase, thus we use the average of optical flux during the
plateau phase (i.e., after the rising part and before the decay
phase) to calculate the $L_{1-7.5}$ values: We average the optical
data of GRB 050820A during the period of $t_{\rm obs}=230-722$s; GRB
060418 of $107-137$s; GRB 060607A of $159-205$s; GRB 080810 of
$67-261$s; and for GRB 100906A we use the peak flux at 115s.

Given $L_{1-7.5}$ and $\Delta t$, the maximum dust destruction
radius $R_d$ can be determined to be as function as density $n_{\rm
H}$. Furthermore, the time $t_b$ that the optical flux rise to the
top value can be estimated from the observed optical light curve.
Once given $f(r=\Delta R)=t_b/\Delta t_{\rm obs}$ and combined with
the condition of $R_d=\Delta R$, one obtains a minimum $n_{\rm H}$
value, $n_{\rm H}>n_{\rm H,0}$, otherwise, the destruction front
cannot reach the edge of the cloud and no optical-UV emission
escapes from the cloud. Finally, we apply eq. (\ref{Luminosity}) to
fit the rising part of the optical light curve by taking $n_{\rm H}$
(in the range of $n_{\rm H} > n_{\rm H,0}$) and $t_b$ as free
parameters. The best fit gives us the resulted values of $n_{\rm H}$
and $\Delta R$.

\begin{figure*}
\includegraphics[angle=0,scale=0.4]{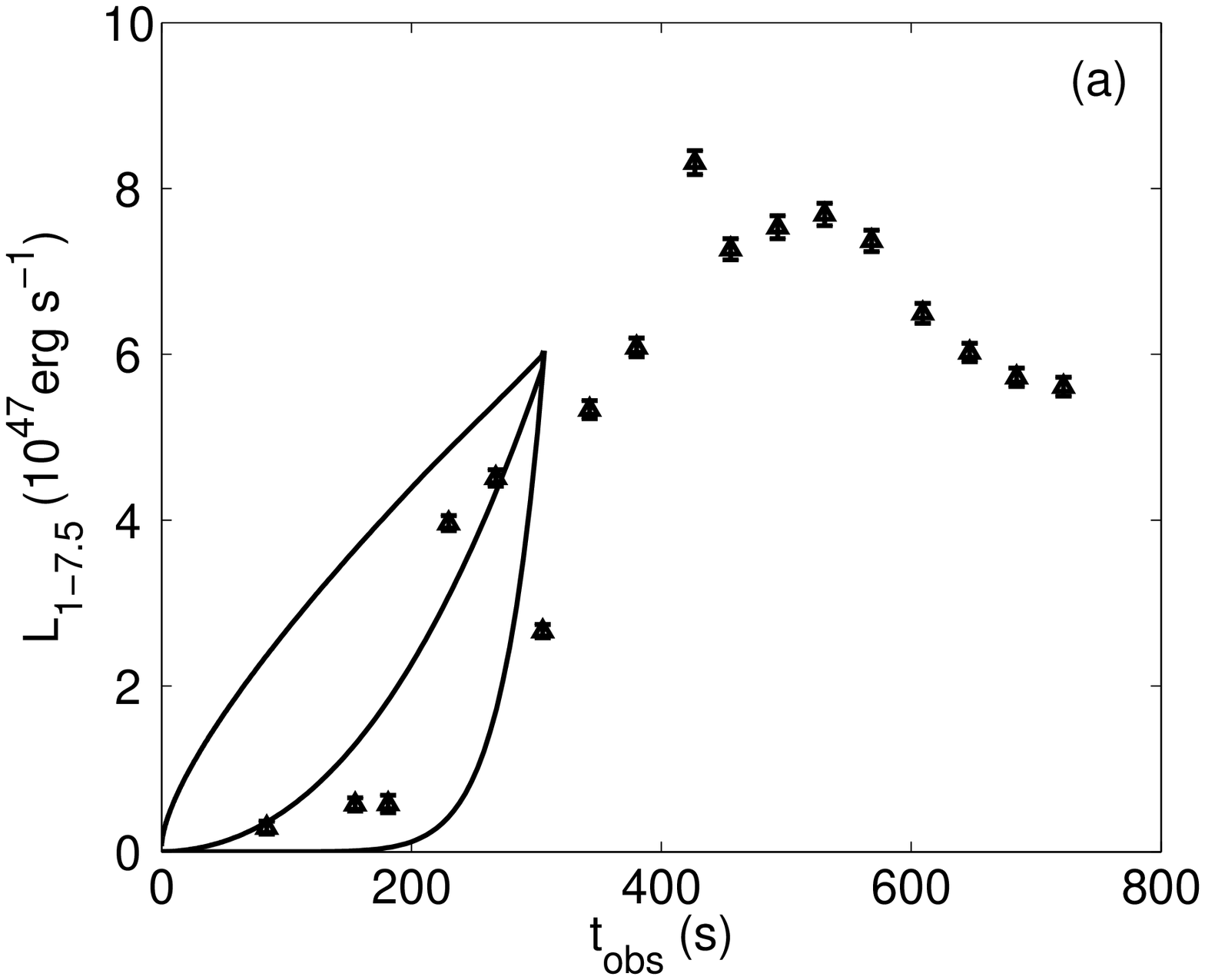}
\includegraphics[angle=0,scale=0.4]{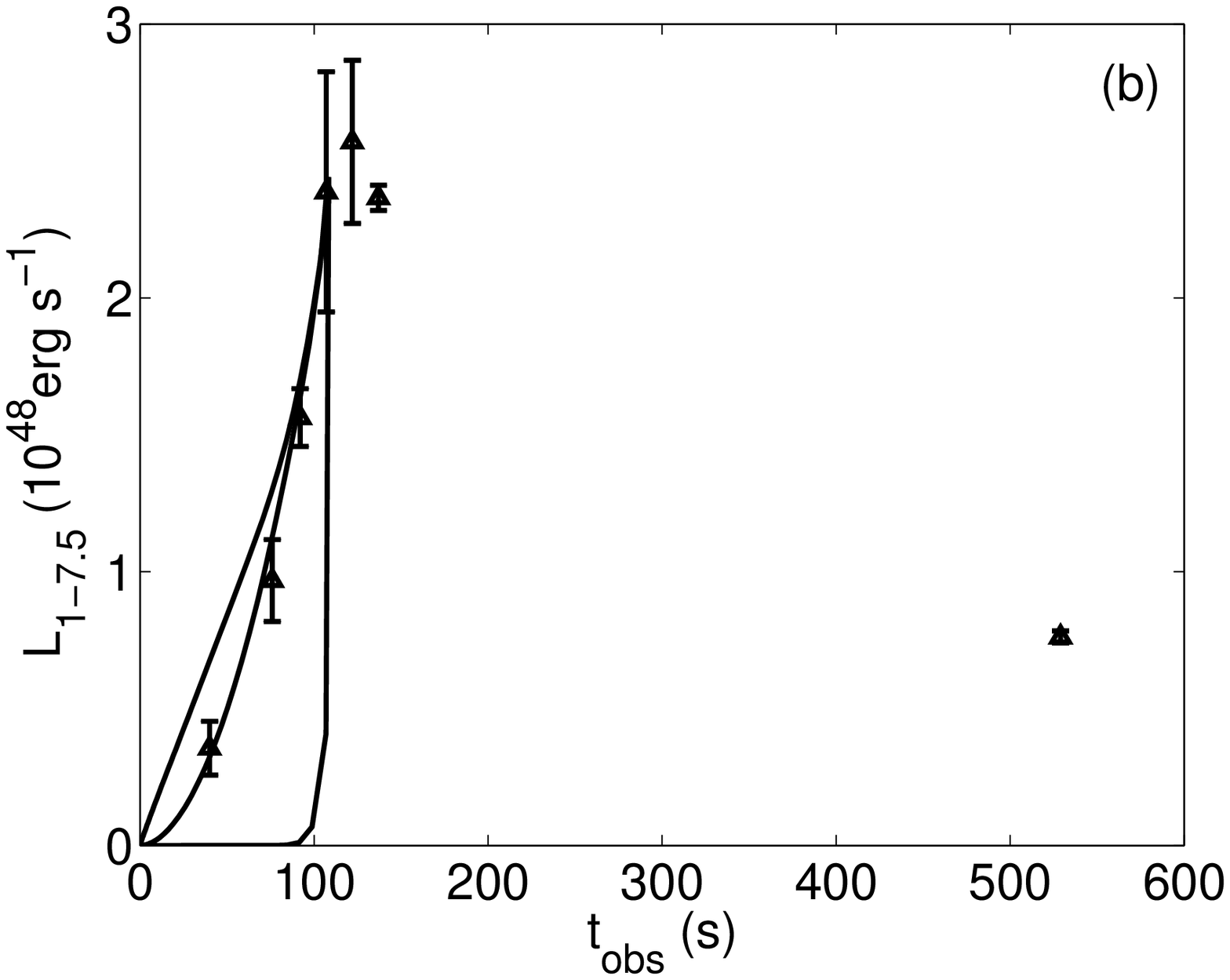}
\includegraphics[angle=0,scale=0.40]{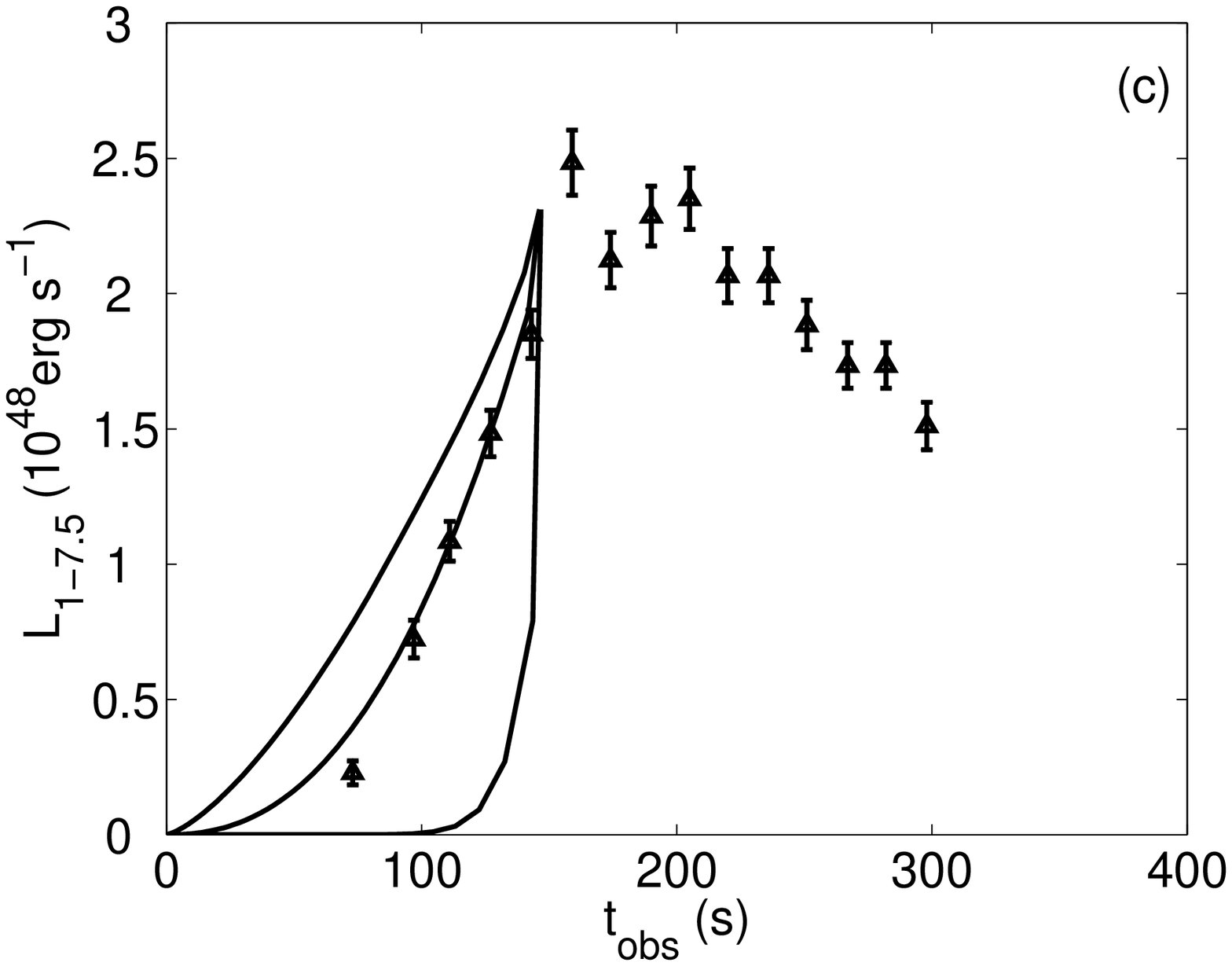}
\includegraphics[angle=0,scale=0.40]{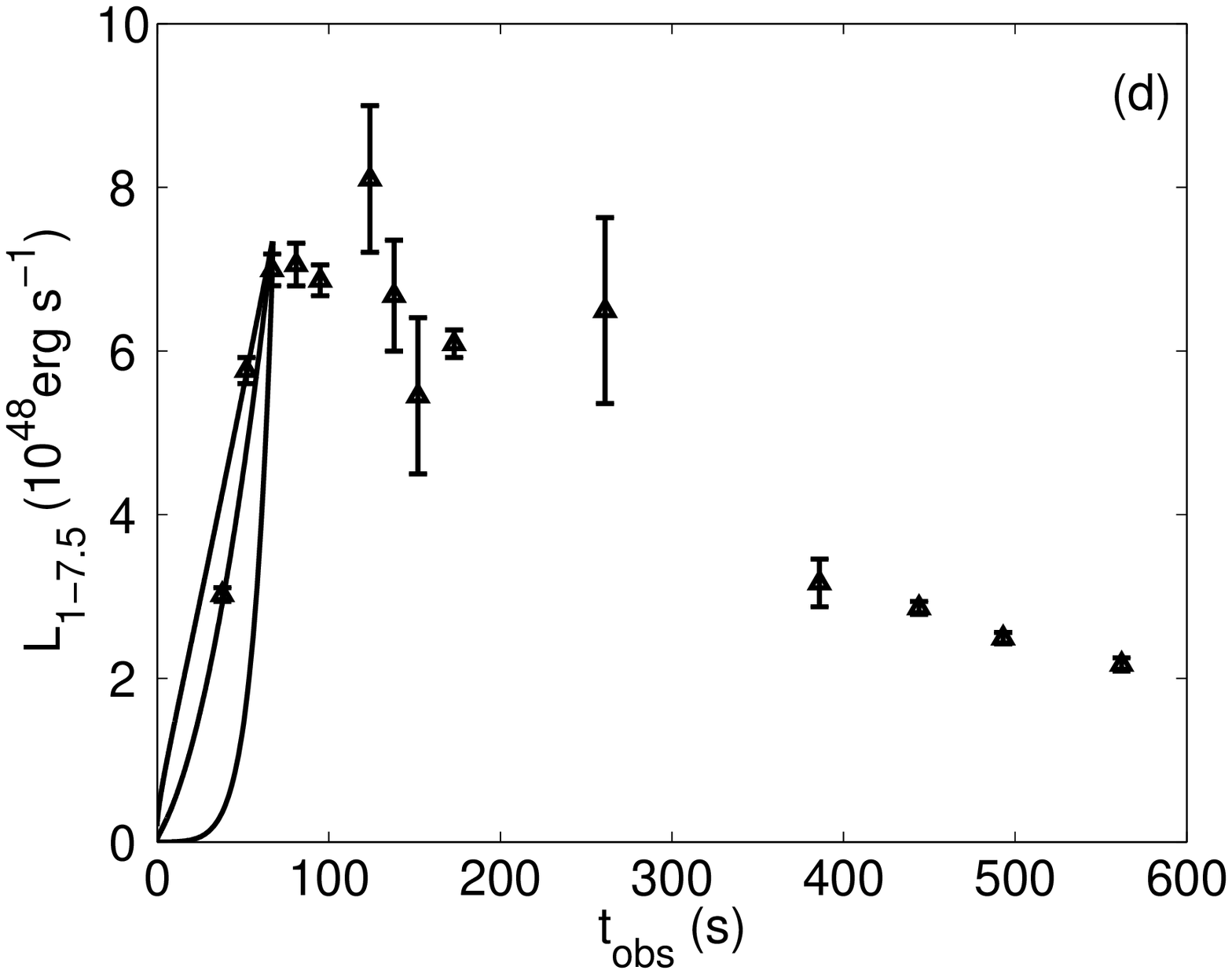}
\includegraphics[angle=0,scale=0.40]{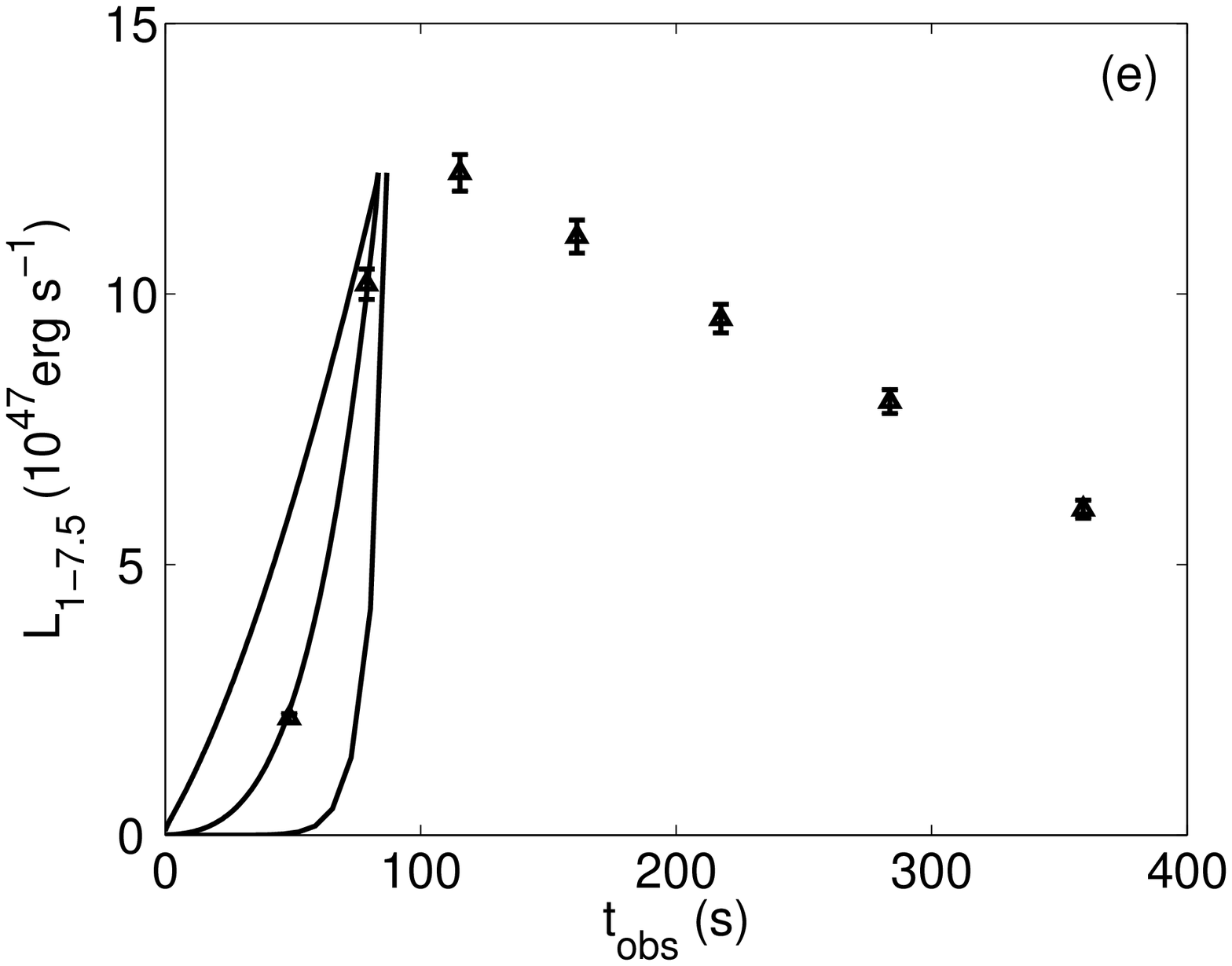}
\hfill \caption{The light curves of prompt optical emission from
four other GRBs in Sample I besides GRB 080319B: GRBs 050820A, 060418,
060607A, 080810 and 100906A. The black triangles are the optical data. The
solid lines are the predictions from radiation-dust interaction
model, with the parameters $n_{\rm H}$ and $t_b$ marked.}
\label{fit1}
\end{figure*}

The resulted values of $n_{\rm H}$ and $\Delta R$ is also listed in
Table \ref{Tab:poptical}. Illustrations of our fitting results for
the four bursts included in Sample I are shown in Fig. \ref{fit1}.
From the fitting results, we find that the density of the
surrounding molecular clouds are in the range of
$10^3-10^4$cm$^{-3}$, while the size, as implicated by $\Delta R$,
is in the order of $\sim10$ pc. However, due to small number of GRBs
with optical rising part detected in prompt emission, it is
impossible to give the statistic discussion of the properties of
local molecular clouds. Furthermore, the observed data points in the
optical rising part are usually sparse for individual GRB, which may
induce large errors in light curve fitting. The future precise
observations are needed to test the model and constrain the
properties of local environment more precisely.

\subsubsection{Sample II}
For GRBs in Sample II, because there is no optical rising detected
due to the delay of optical observations, we cannot well constrain
the properties of the surrounding environments. However we still try
to make some constraints, although rough. In these bursts, the time
$t_b$ when the optical-UV flux reaches the plateau, i.e., when the
dust destruction front reaches the edge of the cloud, $R_f=\Delta
R$, can be considered to be smaller than the start time of the
optical observations, $t_{\rm op}$. Thus we have $t_b<t_{\rm op}$.
Moreover, we take the flux of the first optical data point to
calculate the mean luminosity in optical-UV band. In these bursts,
the maximum dust destruction radius must be within the boundary of
the cloud, $R_d<\Delta R$, otherwise the optical emission from these
GRBs cannot emerge.

\begin{figure*}
\includegraphics[angle=0,scale=1.2]{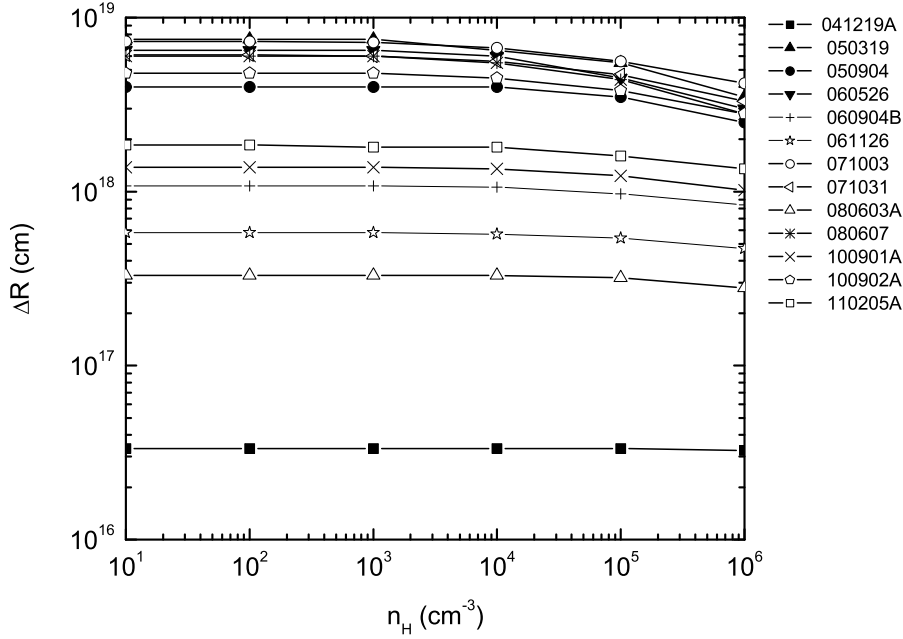}
\hfill \caption{The constraints on the sizes and densities of
molecular clouds around GRBs in Sample II. The allowed parameter
region for each GRB is that below the line corresponding to it. }
\label{fig:sample2}
\end{figure*}

Thus we constrain the properties of the molecular clouds of
sample-II GRBs as follows. Considering an optical-UV flash with
luminosity $L_{1-7.5}$ and duration of $\Delta t=t_{\rm op}/(1+z)$,
we can calculate the maximal dust destruction radius as function of
surrounding density. This puts an upper limit to the value of
$\Delta R$ of the relevant GRB. The results for all GRBs in sample
II are shown in Fig\ref{fig:sample2}. We see that although there is
no good constraints on density $n_{\rm H}$, the value of $\Delta R$
is quite well constrained since $\Delta R$ does not vary much with
$n_{\rm H}$. All except GRB 041219A have upper limits of $\Delta
R<0.1-2$~pc, somewhat less than those of sample-I GRBs. This might
be reasonable since the luminosity of sample-II GRBs are generally
smaller than that of sample-I GRBs. GRB 041219A has exceptionally
small luminosity then its value of $\Delta R$ is smaller than
$\sim0.01$ pc.

\section{Discussion and conclusions}
Long GRBs are believed to be the explosions of massive stars,
therefore the GRBs may occur in the molecular clouds where their
progenitors were born.  We show in this work that the prompt
optical-UV emission from GRBs, if originally emitted simultaneously
with $\gamma$-ray emission, may appear with relative time delay in
observations, due to the dust extinction. This can well explain the
optical delayed onset observed in GRB 080319B, and the number
density and the size of the molecular cloud are roughly constrained
to be $n_{\rm H}\sim10^3$cm$^{-3}$ and $\Delta R\sim 8$pc,
respectively. We also investigate the other GRBs with good
optical-UV data, and find the densities and sizes of the molecular
clouds in the range of $n_{\rm H}\sim10^3-10^4$cm$^{-3}$ and $\Delta
R\sim10$pc.

We use a simple picture that the effects of extinction is
approximated as a narrowing of the optical pulse, retaining a
rectangular profile. This neglects that there may be fluctuation of
the original flux with time, and that the dust destruction front is
not a zero-thickness one. Thus the constraints on the molecular
clouds only make sense by order of magnitude.

The resulted $n_{\rm H}$ and $\Delta R$ constraints suggest high
column densities of gas around GRBs, $\sim10^{22}-10^{23}$cm$^{-3}$.
It is interesting to note that Galama \& Wijers (2001) obtain
similar range of column densities by observations of X-ray afterglow
spectra. Moreover, our constraints are also consistent with those
giant molecular clouds found in Milky Way, which are observed to
have sizes of $10-30$ pc and average gas densities of
$10^2-10^3$cm$^{-3}$ (Winnewisser et al. 1979; Goldsmith 1987).
Although the GRB local molecular clouds from our constraints seem to
be slightly denser, the low statistics of GRBs with prompt optical
detection and the sparse data points for individual GRBs prevent us
from giving clear conclusions.

One may expect that the initial fast rise of the prompt optical flux
can be produced by the afterglow forward shock due to sweep-up the
circumburst medium before deceleration. However, the multi-band
observations of two sample-I GRBs 060418 and 060607A show spectral
index in the optical band of $F_\nu\propto\nu^{-0.9}$ and
$F_\nu\propto\nu^{-0.8}$, respectively. This implies the injection
frequency below the optical band, $\nu_m<\nu_{\rm opt}$, and require
extremely unusual afterglow model parameters, e.g., postshock
electron energy far smaller than typical value,
$\epsilon_e\la10^{-3}$. Moreover, most GRBs in sample I show
up-rising even faster than $F_\nu\propto t^3$ (e.g., GRBs 060418,
060607A, 080319B and 100906A), which is faster than the model
prediction for the pre-deceleration forward shock emission at
$\nu>\nu_m$. One may also expect the up-rising part can be accounted
for by the reverse shock emission due to shock swept-up of outflow
material but the predicted temporal slope is not faster than
$F_\nu\propto t^2$ (Kobayashi 2000). Thus the prompt optical
emission is more likely to be generated within the outflow.

The density of the surrounding from our constraints is higher in
general than the medium density implicated by afterglow modeling. The X-ray
absorber must lie within $1-5$ pc from the GRB thus probing
the innermost region in the close vicinity of the GRB explosion. Comparing the HI column densities from Ly$\alpha$ absorption to the metal column densities from X-ray absorption in GRB afterglows, Watson et al. (2007) found there no correlation
between the column density values, and the X-ray absorptions
often far exceed the HI column densities. Based on a detailed study of the absorption pattern, Campana et al. (2011) found a
high-metallicity absorbing medium for GRB 090618 and a best-fitting column densities $6\times 10^{17}$ cm$^{2}$ in Ne and Si.
However there is no contradiction here because the size of observed
afterglow is usually sub-pc scale, but the region concerned here is
in much larger scale, $\sim10$pc, as resulted from the constraints.
Thus, it may be that in the place very close to the GRB location the
medium density is low while the further-out region has much denser
gas. This is reasonable that the vicinity of the GRB source may be
affected by the progenitor before the GRB explosion.

In our simple model, for given luminosity and duration of the prompt
optical-UV emission, the maximum dust destruction radius can be
determined. Once it is within the boundary of the cloud, $R_d<\Delta
R$, there will be no prompt optical emission observed, neither the
optical afterglow emission. The GRB will appear as being optically
dark in this case. It is interesting to note that only 60\% GRBs
observed by BAT/Swift are detected by UVOT/Swift in the optical
afterglows. The ``dark burst'' (van der Horst et al. 2009) are still
mystery now. If dust extinction is the reason, then by our simple
model, this suggests that the maximum dust destruction radii and the
sizes of the molecular cloud are statistically comparable, i.e.,
$R_d\sim\Delta R$, thus the bright and dark bursts are comparable in
numbers. Indeed, as shown in Table \ref{Tab:poptical}, those bright
GRBs in sample I with better observations and hence better
constraints, $R_d$ and $\Delta R$ values are similar.

There are quite a few small robotic telescopes that have been built
and installed around the world in order to detect the optical
counterparts in the early phase of $\gamma$-ray bursts, such as
Super-LOTIS (Park et al. 1997), TAROT (Klotz et al. 2009), PROMPT
(Reichart et al. 2005), ROTSE-III (RykoR et al. 2009),
SkyNet\footnote{http://skynet.unc.edu/}, WIDGET (Urata et al. 2011),
MASTER\footnote{http://observ.pereplet.ru/}, Pi of the sky (Burd et
al. 2005), and
TORTORA\footnote{http://www.eso.org/public/images/eso0808a/} etc.
With their large Field of View (FOV) and fast slewing abilities,
these telescopes are able to detect the prompt optical emission in
minute timescale after the trigger of GRBs by $\gamma$-ray
detectors. In the case of the ``naked-eye GR'' 080319B (Racusin et
al. 2008), the prompt optical emission was caught by the TORTORA and
Pi of the sky even with zero time delay. Besides, the
UFFO-Pathfinder (Chen et al. 2011), which aims at prompt optical
detection in subsecond timescale, will be launched soon; the
Ground-based Wide-Angle Camera array (GWAC), with a larger field of
view ($\sim$ 8000 square degrees), as a part of ground system of the
Chinese-French SVOM mission (Paul et al. 2011), aiming at search for
the optical emission in zero delay, will be constructed in the near
future. All these robotic telescopes and the planed projects will
make a larger and better sample of prompt optical emission from GRBs
in the future, leading to more precise constraints on the GRB local
environments.

\begin{acknowledgements}
We would like to thank the useful discussions at pulsar group of PKU
and at SVOM group of NAOC. This work is supported by National Basic Research Program of
China-973 Program 2009CB824800, China
Postdoctoral Science Foundation funded project (No. 20110490590),
the National Natural Science Foundation of China (Grant Nos.
111030262) and the Foundation for
the Authors of National Excellent Doctoral Dissertations of China.
\end{acknowledgements}

\label{lastpage}

\end{document}